\documentclass{article}

\usepackage[final]{nips_2017}
\usepackage[utf8]{inputenc} 
\usepackage[T1]{fontenc}    
\usepackage{hyperref}       
\usepackage{url}            
\usepackage{booktabs}       
\usepackage{amsfonts}       
\usepackage{nicefrac}       
\usepackage{microtype}      

\usepackage{etoolbox}

\usepackage{graphicx}      
\graphicspath{ {Figures/} }

\title{ProLanGO: Protein Function Prediction Using Neural~Machine Translation Based on a Recurrent Neural Network}

\author{
  Renzhi Cao \\
  Department of Computer Science\\
  Pacific Lutheran University\\
  Tacoma, WA 98447 \\
  \texttt{caora@plu.edu} \\
  \And
  Colton Freitas \\
  Department of Computer Science\\
  Pacific Lutheran University\\
  Tacoma, WA 98447 \\
  \texttt{freitacr@plu.edu} \\
  \And
  Leong Chan \\
  School of Business\\
  Pacific Lutheran University\\
  Tacoma, WA 98447 \\
  \texttt{chanla@plu.edu} \\
  \And
  Miao Sun \\
  Baidu Inc.\\
  1195 Bordeaux Dr, Sunnyvale, CA 94089, USA \\
  \texttt{miaosunwork@gmail.com} \\
  \And
  Haiqing Jiang \\
  Hiretual Inc.\\
  San Jose,  CA 95131, USA \\
  \texttt{stevenjiang@hiretual.com} \\
  \And
  Zhangxin Chen \\
  School of electronic engineering\\
  University of Electronic Science and Technology of China\\ Chengdu, 610051, China \\
  \texttt{zhangxinchen@uestc.edu.cn} \\
}

\begin{document}

\maketitle

\begin{abstract}
With the development of next generation sequencing techniques, it is fast and cheap to determine protein sequences but relatively slow and expensive to extract useful information from protein sequences because of limitations of traditional biological experimental techniques. Protein function prediction has been a long standing challenge to fill the gap between the huge amount of protein sequences and the known function. In this paper, we propose a novel method to convert the protein function problem into a language translation problem by the new proposed protein sequence language “ProLan” to the protein function language “GOLan”, and build a neural machine translation model based on recurrent neural networks to translate “ProLan” language to “GOLan” language. We blindly tested our method by attending the latest third Critical Assessment of Function Annotation (CAFA 3) in 2016, and also evaluate the performance of our methods on selected proteins whose function was released after CAFA competition. The good performance on the training and testing datasets demonstrates that our new proposed method is a promising direction for protein function prediction. In summary, we first time propose a method which converts the protein function prediction problem to a language translation problem and applies a neural machine translation model for protein function prediction.

\end{abstract}
\section{Introduction}

With the wide application of next generation sequencing techniques, it takes a short time with low cost to generate millions of protein sequences \cite{1}. Understanding the properties (e.g., protein function, protein structure) of each protein sequence becomes an urgent task, which not only helps us better understanding their role in our life, but more importantly their potential biomedical and pharmaceutical applications, such as drug discovery \cite{2,3}. The traditional biological experimental methods to determine a protein’s properties (e.g., protein function and structure) can be very slow and also resource-demanding \cite{3,4}. Moreover, sometimes it has the limitation that may not faithfully reflect the protein’s activity in vivo \cite{3,5}. Due to these limitations, the computation method that could accurately and quickly predict protein function from its sequence is greatly desired. The computation protein function prediction method can be used to fill the gap between the large amount of sequence data and the unknown properties of these proteins.

Protein function prediction is usually treated as a multi-label classification problem. Researchers have tried different computation methods in the last few decades for this problem \cite{6,7,8,9,10,11,121}. In general, the following methods are used for protein function prediction.

The first and the most widely used method for protein function prediction is the Basic Local Alignment Search Tool (BLAST) \cite{12} used to search the query sequence against the existing protein databases, which contain experimentally determined protein function information, and then use these homologous proteins’ function  information for the function prediction of the query sequence. For~example, the~methods of Gotcha \cite{13}, OntoBlast \cite{14}, and Goblet \cite{15}. Except for using the BLAST tool, some methods use the tool PSI-BLAST \cite{12} 
 to find the remote homologous, such as the PFP 
 method~\cite{16}.

The second includes network based methods. Most methods in this category use protein--protein interaction networks for protein function prediction based on the assumption that interacted proteins share similar functions \cite{10,17,18,19,20,21,123}. Instead of protein--protein interaction networks, some methods use other kind of networks for protein function prediction, such as gene--gene interaction network and domain co-occurance networks \cite{3,10,22}.

The third is other information based methods. Some methods use protein structure or microarray gene expression data \cite{23,24,25,26} or combinations of different resources for protein function prediction \cite{27,28,29,30}.

The most challenging method is to directly use protein sequence without searching any database or any other resource for protein function prediction. For this kind of methods, the machine learning techniques are usually used to predict protein function from scratch. Currently most machine learning methods use features generated from the protein sequences for model training, and use that model to classify a number of function categories for an input sequence \cite{104,105}. The most commonly used features in these methods are protein sequence \cite{120,122}, protein secondary structure, hydrophobicity, subcellular location, solvent accessibility, etc. For example, PANNZER 
and MS-kNN 
use the machine learning method k-nearest neighbour to predict protein function \cite{31}. SMISS 
 uses an apriori algorithm to mine the association rules for protein function prediction \cite{3}, whereas methods like SVMProt use the machine learning method SVM 
 for protein function prediction \cite{33,34,35}, and some methods use naïve Bayes classifiers~ \cite{36,37}. Except for the protein sequence, most other features are usually predicted from the protein sequence, so more errors could be involved for these features and also for the final protein function prediction.

The latest state-of-the-art machine learning method---deep learning which uses multiple layers representation and abstraction of data has drastically improved the accuracy of speech-recognition, visual object recognition, language translation, protein structure prediction, etc. \cite{102,103,38,39,40,41}. For example, Google developed its neural machine translation system based on Neural Machine Translation (NMT) in November 2016 to increase accuracy in Google Translate \cite{42}. It would be interesting to apply these latest machine learning methods for the protein function prediction problem. Based on our knowledge, currently there are very few methods that only use protein sequence without any additional features. Researchers have tried to derive more features from protein sequence, and applied machine learning techniques for the protein function prediction problem \cite{110}, but most methods have additional features combined with protein sequence to prediction protein functions \cite{31,3}. The reason is because protein function prediction from sequence only is really challenging, there is no such kind of method that achieves or is even close to the state-of-the-art performance. In addition, there is no method that applies the state-of-the-art machine learning method Neural Machine Translation (NMT) for the protein function prediction problem. In this work, we developed a novel method called ProLanGO, which first converts the protein sequence and protein function into the “ProLan” and “GOLan” language, respectively, and then builds the NMT model based on recurrent neural network to translate “ProLan” to “GOLan” for the protein function prediction. Our new proposed method based on the ProLanGO model provides a new way of predicting protein functions.

Despite the protein function prediction methods, how to test these methods is also important. There is a worldwide experiment---the Critical Assessment of Function Annotation (CAFA,\linebreak \url{http://biofunctionprediction.org/}) \cite{2,43}, which is designed to provide a large-scale unbiased benchmarking of different protein function prediction methods to test their ability to predict protein function on the same set of proteins within a specific time frame \cite{2,43}. There are three stages for CAFA: prediction phase, the organizer releases protein sequences with unknown or incomplete function for collecting the submission of predictions (usually about four months); target accumulation phase, waiting for the biological experiment to determine protein’s function (usually between six and 12 months); and analysis phase, the organizer evaluates and ranks the performance of different methods based on their prediction on proteins whose function is known during the target accumulation phase. Until now, three CAFA experiments have been conducted. The first CAFA (CAFA1) showed the good performance of applying a machine learning method to integrate multiple sequence hits and data types for protein function prediction. We have attended the third CAFA (CAFA3) in 2016 to blindly test our method on a large scale of protein sequences to evaluate the performance of our model.

The rest of the paper is organized as follows. In Section 2, we describe some results on training and testing datasets. In Section 3, we describe the data and detailed method to build and test our method. In Section 4, we summarize our work and discuss the future directions.

\section{Method}
In this paper, we construct the method ProLanGO which uses a neural machine translation model based on a recurrent neural network to predict protein function from protein sequence, and also we construct a baseline method which uses probability theory as a comparison.
\subsection{Data Preparation}
The training data of our model comes from UniProtKB knowledge database \cite{44} with the version of 1 June, 2016. We only consider Gene Ontology (GO) terms ID which are valid from the date of 1 June, 2016. In total, 523,990 protein sequences and 42,819 GO terms are used to train our model. We blindly test our method on CAFA3, which includes 130,787 protein sequences in 23 species.

\subsection{Understanding Protein Sequence as a Language}
Our assumption is that protein sequences are similar to human language; each protein sequence can be represented as a sentence consisting of protein “words”.  Proteins are composed of amino acids, and we use 20 letters to represent each of 20 standard amino acids found in protein. Each protein sequence  is a string of these 20 letters, and we use to represent the set of these 20 letters.  We let $A^k$ be k-mers or fragment with length k, which is a k-th cartesian power of $A$ or a string of k letters from $A$. The k-mers are considered as protein “word” \cite{106,107,108,109}. The following algorithm has been used to divide protein sequence into a set of k-mers or protein “word”:

Step 1. Scan the whole training dataset - UniProtKB knowledge database to get a protein “word” database, which includes all k-mers whose frequency f$^k$ is larger than 1,000, where  k $\in$ [3,5].

Step 2. For each protein sequence $s$, we divide it into a set of protein “word” based on Step 1. In this step, the longer protein “word” are preferred for matching, and each amino acid can only belong to one protein “word”. If t continuous amino acids cannot match with any fragment from Step 1, we let these continuous amino acids be a new protein “word” A$^t$, t $\in$ [1,$\infty$).
\vspace{6pt}

We ranked all k-mers based on its frequency in training dataset, and the frequency of k-mers ranked at $1000^{th}$ is 47,802, 7167, 1232, 650, 565 for each k of three, four, five, six, and seven respectively. We did not evaluate performance of k-mers with k larger than five, because low frequency in training dataset for these large k could be less reliable. We have also tried to add six-mers and seven-mers (the best loss on testing dataset is 8.62, 9.26, 8.59 for $k$ $\in$ [3--5], [3--6], and [3--7] respectively), and no significant improvement has been found in our experiment. More details of our experiments are in Supplementary {Tables S1--S3}.

\subsection{Understanding Gene Ontology Terms as a Language}
The traditional protein function prediction methods directly predict Gene Ontology (GO) terms for a protein sequence. The GO terms are represented by GO term ID using a seven digit number (e.g., GO:0000001), and these GO terms are used to form a directed acyclic tree structure based on the relationship between each other (e.g., some GO terms are part-of other GO terms). There are three trees to represent the GO terms in each of the three biological categories (Biology Process, Cellular Component, Molecular Function). The traditional GO term ID does not considered the relationship between each GO term (e.g., GO:0015772 is a GO:2001088, but it is not reflected from the GO ID). Moreover, it needs seven digits to describe a GO term. In this paper, we describe a new way to encode GO terms, and consider the encoded GO terms as a new language to describe functions of proteins.

We do a depth-first search (DFS) on Biological Process, Cellular Component, and Molecular Function tree respectively, and assign the 26-base Alphabet number to each GO term as a new ID (Alphabet ID). For example, GO:0008150 is the root of Biological Process (BP) tree, and there are 28,678 GO terms in BP tree, so GO:0008150 is the number 28,678th node to be visited by DFS, and we can convert 10-base number 28,678 to 26-base Alphabet ID: BQKZ. In theory, a four letter Alphabet ID can represent $26^4$ GO terms, which is more than the current total number of GO terms.  In summary, the new Alphabet ID has a maximum length of four, and considers the relationship of GO terms in the Gene Ontology tree. For a protein with multiple functions (GO terms), we simply convert each GO term to a Alphabet ID, and the order of these Alphabet IDs  is  based on the order of the respective GO term from the UniProtKB knowledge database (See Figure ~\ref{figure1}).

\subsection{ProLanGO Model by Neural Machine Translation Based on Recurrent Neural Network}
\begin{figure}[h]
\centering
\includegraphics[width=14cm]{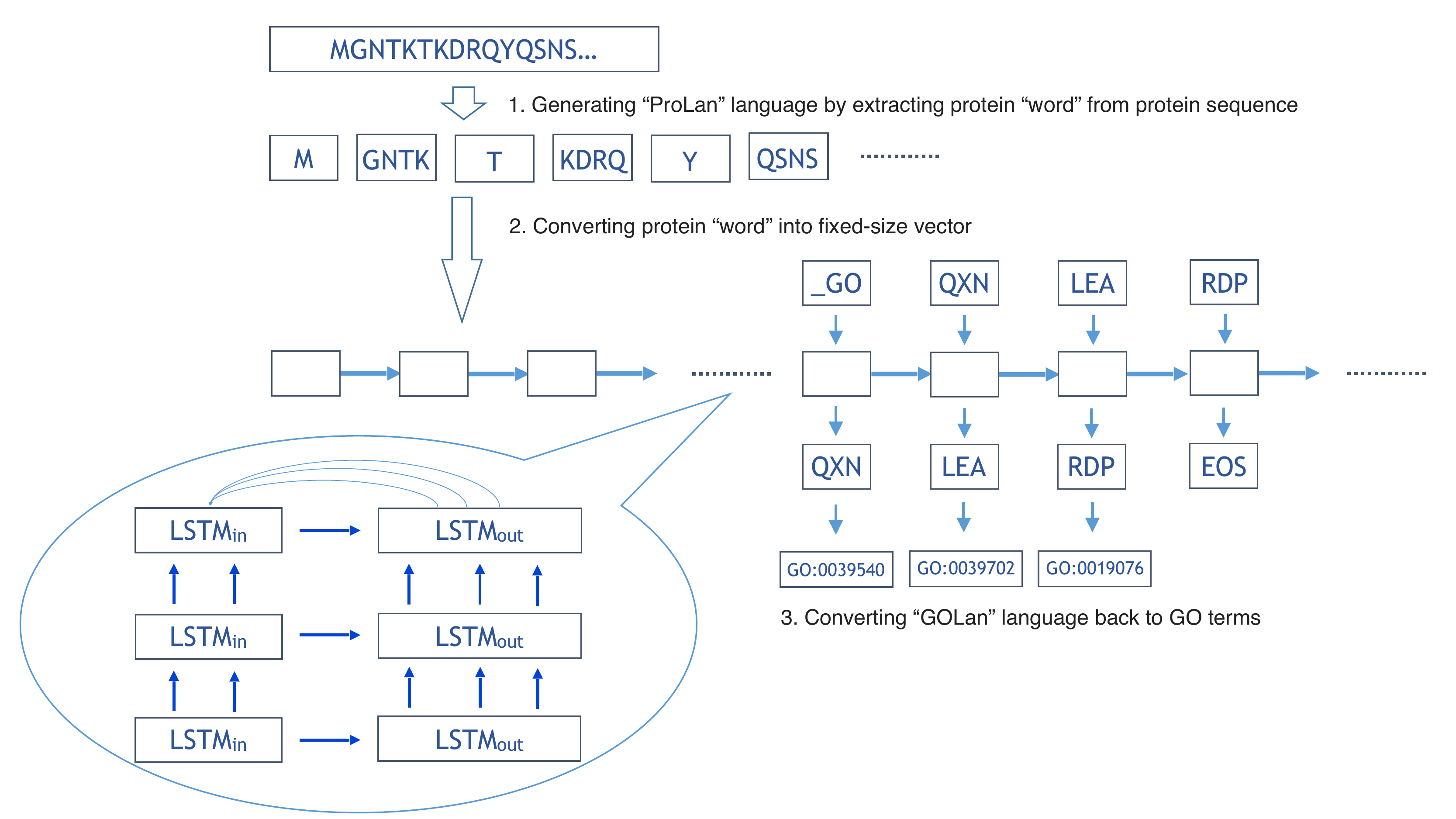}
\caption{The flowchart of our method for protein function prediction. Three layers of RNN 
	is used for the NMT 
	model.}
\label{figure1}
\end{figure}

In the previous sections, we convert protein sequences into a new language based on the protein “word” generated from UniProtKB database; we call it the “ProLan” language. Also, the functions of each protein are converted into a new language based on the Alphabet ID, we let it be the “GOLan” language (See Figure ~\ref{figure1}). We now convert the protein function prediction problem into a language translation problem. The neural machine translation model based on two recurrent neural networks is used to solve the language translation problem \cite{45,46}. The first recurrent neural network encodes a sentence in “ProLan” into vectors of fixed length, and the second recurrent neural network decodes the representation into a sentence of “GOLan”. The encoder and decoder RNN are trained together to maximize the conditional probability of “GOLan” sentence given “ProLan” sentence. In~general, there are two multi-label classification methods \cite{113}: problem transformation methods and algorithm adaptation methods. Our method based on RNN is one type of problem transformation method. The RNN could be considered as a chain of repeating modules of a neural network, and each neural network is used to handle a single-label classification problem by outputting a “word” in a “GOLan” sentence. In another word, our method is suitable for protein function prediction (multi-label classification problem) by transforming it into several single-label classification problems and solving each of them with a module of a neural network in RNN. In addition, the attention mechanism~\cite{47} is used in the system to align and translate jointly, which would be helpful to solve the problem when the “ProLan” sentence is relatively long, because it would be difficult for encoder RNN to convert a long “ProLan” sentence into a fixed length of vector.

There are in total 420,127 different protein “words” in “ProLan”, and 25,160 different Alphabet IDs in “GOLan”. In order to simplify the training and decoding complexity with this big “ProLan” and “GOLan” data, we use sampled softmax \cite{48} based on importance sampling which samples on a small subset of the target “GOLan” Alphabet ID for decoding. In addition, the bucketing and padding technique is used to efficiently handle “ProLan” and “GOLan” with different length L1 and L2. We use a number of buckets and a special symbol PAD to pad each “ProLan” and “GOLan” into the buckets. The following buckets have been used:
\vspace{6pt}

buckets = [(64,3), (109,4), (164,6), (300,10)].
\vspace{6pt}

The buckets are used to improve the efficiency of RNN. For example, let’s assume there are three protein “words” in a ProLan sentence and four outputs in a GOLan sentence, then the length for input (L1) is three, and the length for output (L2) is four. Even though the input length L1 can fit the first bucket (64, 3), the output length L2 is larger than the output limit three for the first bucket. In this case, we finally use the second bucket (109,4), and add special symbol PAD to fill the rest so that the input length L1 would be 109, and output length L2 would be four. Figure~\ref{figure1} illustrates the overall flowchart of our method. In~order to decide the upper and lower bounds for buckets, we analyze the number of input “words” and outputs from ProLan sentences and GOLan sentences in the training data sets. This bounds number is selected so that there are a similar number of data falling into each bucket for training data~sets.

It should be mentioned here that the maximum number of outputs is 10, based on the bucketing and padding technique, which means we can predict a maximum of 10 GO term functions for each protein sequence. Some proteins may contain more than 10 functions and in order to solve this problem, we add an extended NMT model. For this extended NMT model, we still keep the same “ProLan” language, and change a little bit of the “GOLan” language. For the GO terms of the protein in the training dataset, we add all descendants of each of the GO terms in the directed acyclic tree to extend the sentence in “GOLan” language. The same NMT framework is used except the following buckets are used instead:
\vspace{6pt}

buckets = [(64,25),(164,50),(250,60)]
\vspace{6pt}

In this extended NMT model, the maximum number of function prediction could be 60, which is big enough for most proteins. As we could imagine, the new added descendants could be less accurate especially when the descendant is much deeper from the real GO term in the tree structure. A decreased rate is added to each new added descendant GO term, so that the one far away from the original true GO term would have lower score.

Finally, the ProLanGO model is a combination of the NMT and extended NMT model by giving different weights for each model based on their performance on training data.

The prediction part is straightforward, for each protein sequence, we first convert it to a “ProLan” sentence, and then applied the trained neural machine translation model to translate the “ProLan” sentences  to “GOLan” sentences and extended “GOLan” sentences. Finally, we apply the ProLanGO model to combine these predicted “GOLan” sentences and convert these “GOLan” sentences back to GO terms by a simple mapping.

\subsection{Baseline Method on Probability Theory}
In order to benchmark our neural machine translation model for protein function prediction, we create a benchmark method which applies simple probability theory to make the protein function prediction. Let x be the fragments vectors from the “ProLan” space, and y be the Alphabet ID for all possible protein functions. Here, x is a vector [x$_1$,x$_2$,...,x$_n$], and the n is decided by the total number of fragments generated for each protein sequence. Here, y is a vector [y$_1$, y$_2$, …, y$_m$] and m is the total number GO terms. We use the following probability theory to calculate the probability of each y$_j$ given the x generated for each protein sequence:
\vspace{6pt}

$P(x_1,x_2,...,x_n) = 1 - \prod_{i=1}^{n} (1-P(y_j|x_i))$
\vspace{6pt}

P(y$_j$|x$_i$) is calculated from the UniProtKB knowledge database by counting the frequency of y$_j$ in all proteins with x$_i$

\subsection{Evaluation Method}
We evaluate the performance of different methods based on precision and recall of top n predictions ranked by the confidence score of predictions. Here the n is in the range of 1 and 10. The~precision and recall are calculated by the following formula:
\vspace{6pt}

$Precision = \frac{total number of corrected predictions}{total number of GO predictions}$
\vspace{6pt}

$Recall = \frac{total number of corrected predictions}{total number of true GO terms}$
\vspace{6pt}

The true GO terms are determined by biological experiment for each protein and here we simply use the UnitProtKB database to extract the true GO terms for each protein. We propagate all true and predicted GO terms to the root of Gene Ontology Directed Acyclic tree structure, and any propagated GO term predictions that exist in the path of true GO terms to the root are considered to be correct. The~top n metric is used for the evaluation, which pick n GO term predictions based on the confidence score to calculate the precision and recall.

\section{Results}
\subsection{Performance during Training of Neural Machine Translation Model}
\begin{figure}[h]
\centering
\includegraphics[width=14cm]{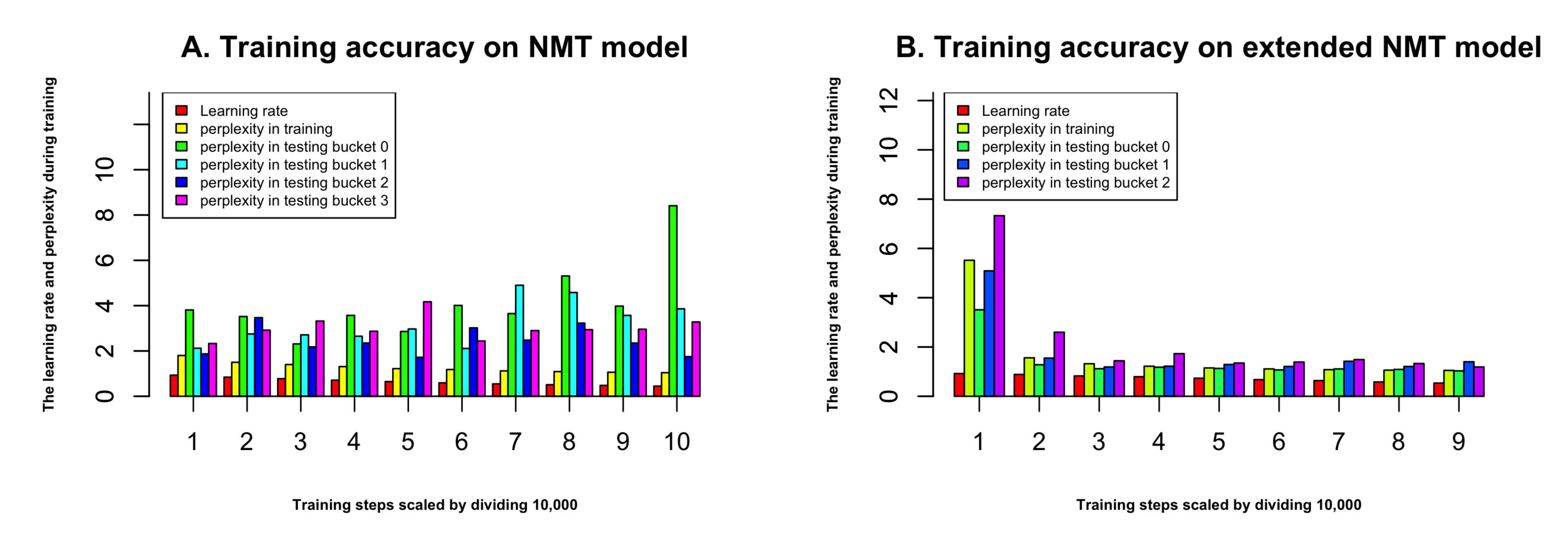}
\caption{(A). The accuracy during training process for NMT model. (B). The accuracy during training process for extended NMT model. The x-axis represents the steps during the training, we scaled it by dividing 10,000, and the first 10,000 steps are not shown in the figure since the perplexity is very big.} 
\label{figure2}
\end{figure}

The overall flowchart of our method is shown in Figure~\ref{figure1}. In the training step, we use 419,192 protein sequences to train the ProLanGO model and another 104,798 protein sequences as testing data to evaluate the performance of our model. The perplexity has been used to evaluate the performance of our language model ProLanGO on training datasets and testing datasets. Figure~\ref{figure2} shows how the perplexity changes on training datasets and performance on testing datasets based on buckets. The~buckets are techniques used in RNN 
to efficiently handle protein sequences with different length, which may not influence the performance too much. For our method, we select the bucket size based on our analysis of protein sequences in the protein database so that it could handle most protein sequences efficiently. We ran 300,000 steps for training, and the learning rate in the range of [0--1]. As~we can see from the picture, the perplexity on training datasets is stable after 10,000 steps, and the perplexity for each buckets on testing datasets is still fluctuating. The result of first 10,000 steps is not included because RNN is not stable with few training steps and the perplexity is huge. Based on the performance for every 200 steps, we finally select the model at step 78,400 (learning rate is 0.5472) and 95,200 (learning rate is 0.5639) for the normal and extended neural machine translation model respectively, and combine these two models as our final ProLanGO model for prediction.

\subsection{Performance on Validation Dataset}
We use the UnitProtKB database on 6/1/2016 to train and test our model and blindly tested on around 100,000 protein sequences from CAFA3. Some of these protein sequences’ function has been released later in UnitProtKB database, and we evaluate the performance of our model on those proteins. We have not performed a homology search against the database, so part of the proteins may have homologies in the database of 6/1/2016, and it may influence the evaluation. However, this problem would be solved after CAFA officially release their evaluation later. In total, we find 332 proteins that are added after 6/1/2016. We propagate both real and predicted GO 
terms to the root of the Gene Ontology Directed Acyclic tree structure as described in method section, and the evaluation of precision and recall is described in method section.

\begin{figure}[h]
\centering
\includegraphics[width=12cm]{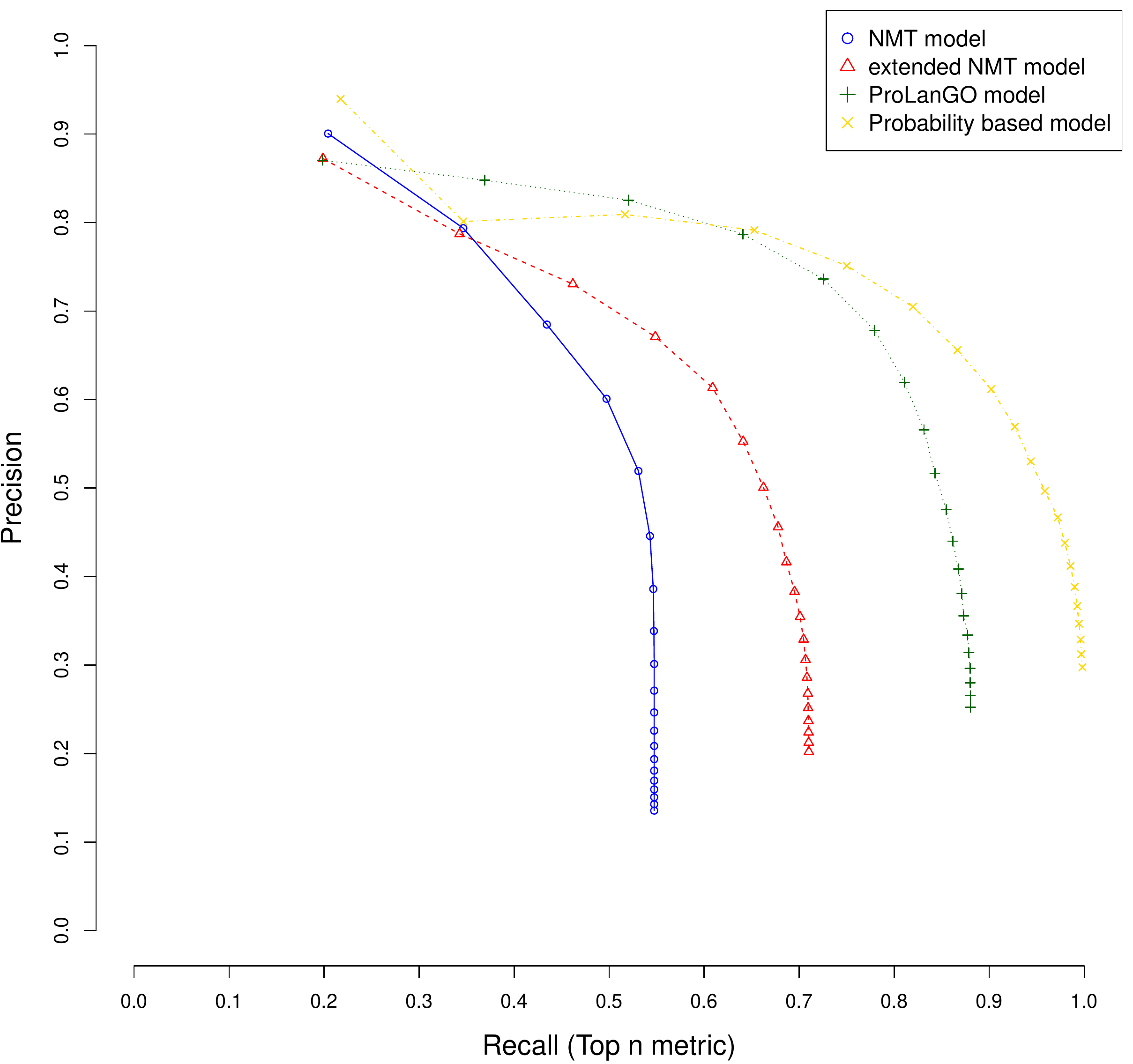}
\caption{Precision and recall of different methods on selected proteins. The top n metric is used, and n is set to be 20. The area under curve (AUC) for NMT, extended NMT, ProLanGO, and Probability based model is 0.286, 0.305, 0.333, and 0.390, respectively.}
\label{figure3}
\end{figure}

Figure~\ref{figure3} demonstrates that the NMT model achieves the relatively lower recall, and the probabilistic based model achieves relatively higher recall. The reason is because NMT model makes six predictions for most proteins, and the maximum number of predictions is 10, however, the probabilistic based model makes more predictions than NMT. The extended NMT model gets relatively higher performance compared to NMT model. The ProLanGO model, which is a combination of NMT and extended NMT model, achieves better performance on both recall and precision. This result shows that our ProLanGO model has potential to accurately predict protein function from sequences.

\subsection{Comparison with Other Selected Methods on CAFA3}

\begin{figure}[h]
\centering
\includegraphics[width=9cm]{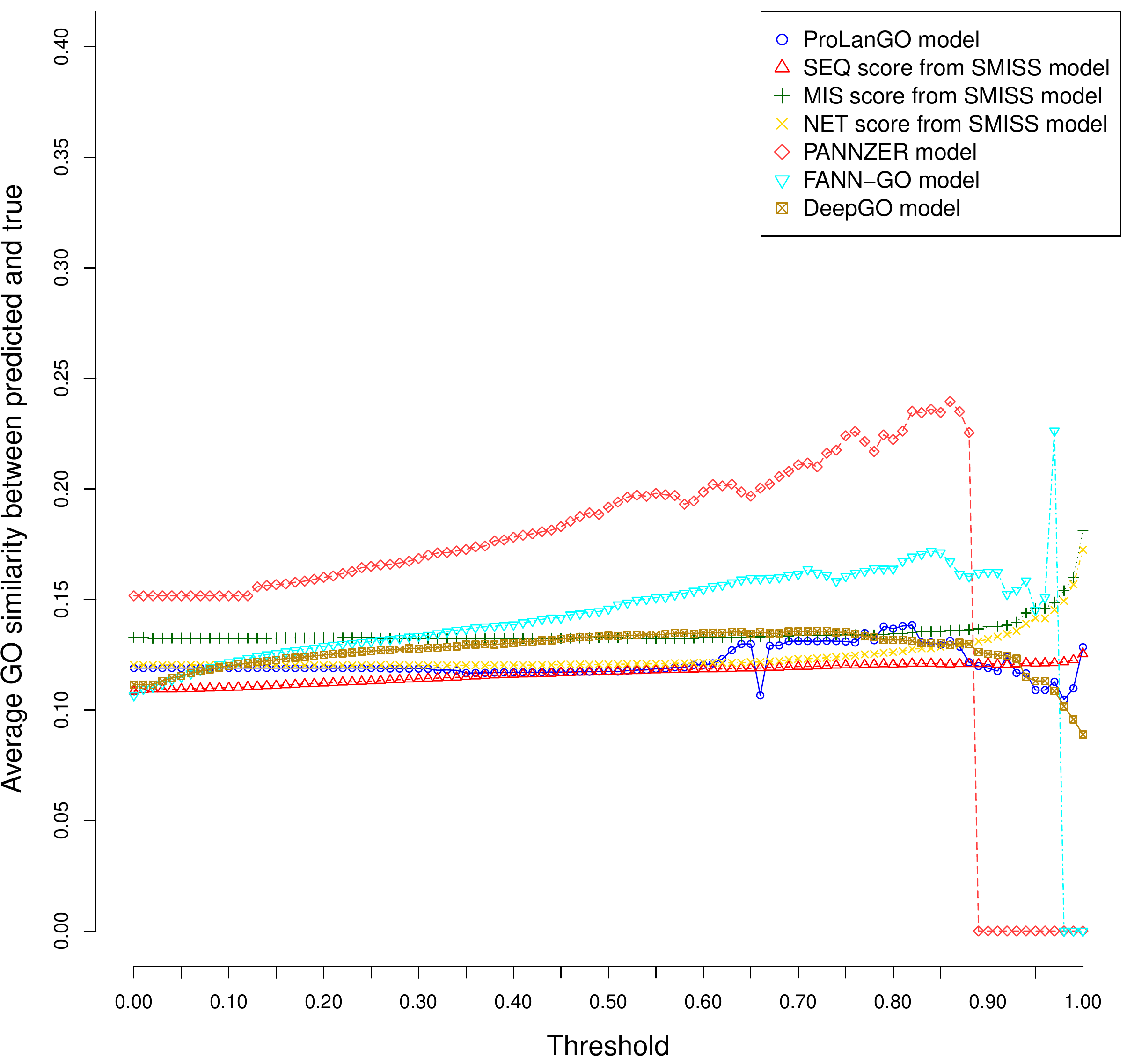}
\caption{Average GO 
	similarities of ProLanGO and other methods on threshold metric.}
\label{figure4}
\end{figure}

\begin{figure}[h]
\centering
\includegraphics[width=9cm]{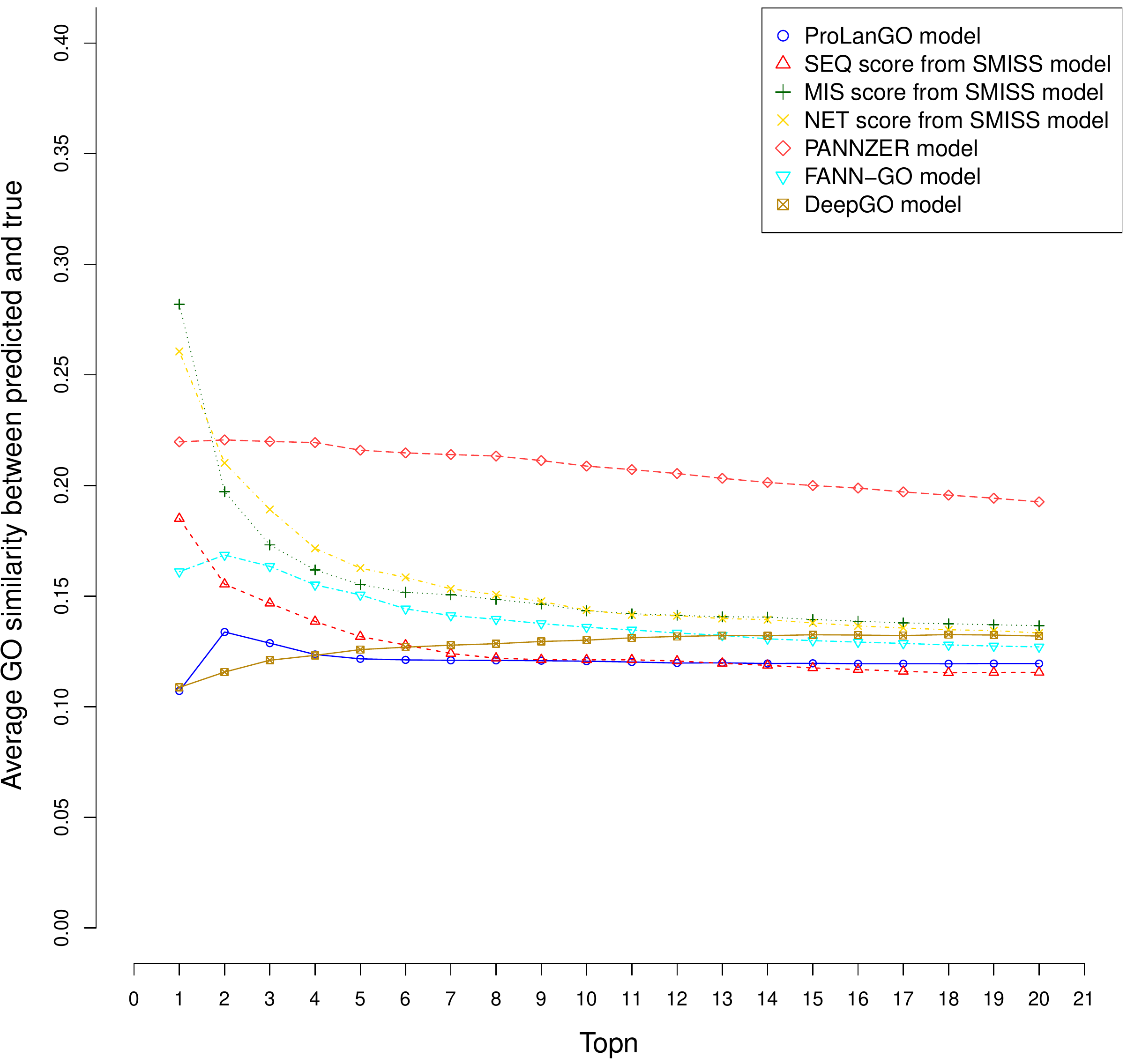}
\caption{Average GO similarities of ProLanGO and other methods on top n metric.}
\label{figure5}
\end{figure}

CAFA3 officially releases true functions for part of CAFA3 targets around 06/05/2017. In total, 674 targets’ BPO  (Biological Process Ontology) function, 539 targets’ CCO (Cellular Component Ontology), and 684 target’s MFO (Molecular Function Ontology) function are released. In order to evaluate the effectiveness of our ProLanGO model in CAFA3 blind experiment, we compare the performance with four state-of-the-art protein function prediction methods in CAFA2: the SMISS model \cite{3,100}, PANNZER model \cite{101}, FANN-GO model \cite{111}, and~DeepGO model \cite{112}. The FANN-GO and DeepGO model are both sequence-based protein function prediction methods, PANNZER is a k-nearest neighbour method, and~the SMISS model contains both sequence-based and network-based predictions. We calculate the average GO similarities of all GO pairs between predicted and true GO for each targets. The GO similarity is defined as the number of shared GO inner nodes divided by the largest number of GO inner nodes when propagating predicted and true GO term in the Gene Ontology Directed Acyclic tree. Figures~\ref{figure4} and \ref{figure5} describe the performance of the ProLanGO model and three different scores from the SMISS model, and~also PANNZER model. Most protein function prediction methods search against protein function databases, and~very few of them try to predict protein function from sequence only without using additional database information. The SEQ 
 score from SMISS model is only based on protein sequence, and~the NET
  score is based on protein--protein and gene--gene interaction networks information. From this figure, we can see that our model achieves a better performance, on average, compared to the SEQ score from SMISS for most of the different thresholds. At the same time, I also want to mention that the gap between sequence based protein function prediction and methods using other information is still big. The MIS 
  score from the SMISS model only uses information from PSI-BLAST searches, and we can find out that the MIS score and PANNZER have better performance comparing to our method. We mention that the average GO similarity score between predicted and true for PANNZER is around 0 for a threshold more than 0.9---the reason could be that our target protein sequences from CAFA are not included in protein sequence databases at the time of our submission. This issue would be solved later when these protein sequences are annotated and added to the database. The result could be different than what we have got after PANNZER updates their database. Our method achieves a similar performance based on Threshold metrics compared to the DeepGO model, while DeepGO performs better than our method for thresholds between 0.10 and 0.80. The FANN-GO model performs better than our method and DeepGO for threshold larger than 0.10. A similar pattern has been observed based on Top n metrics in Figure~\ref{figure5}. However, our model doesn't do especially well for the top one GO term prediction. This is probably because of the lack of ranking GO terms predictions in RNN output. Some additional information might be used in future to improve the ranking of GO terms in our model. Overall, our model shows better performance than the SEQ score in SMISS model, but it also shows that our method is not performing well for the  prediction of top GO terms, which need to be improved in future. Supplementary {Table S4} describes additional resources for protein function prediction methods that we have tried but not succeed because of the availability of these methods.

\section{Conclusions}
In this paper, we propose a novel language model ProLanGO for the protein function prediction problem. We first convert protein sequences into a language space “ProGO” based on the frequency of k-mers on around 500,000 protein sequences, and also encode Gene Ontology terms into a language space “LanGO”. It uses at most four letters to represent a GO term, while the relationship of these GO terms in Gene Ontology Directed Acyclic Graph is considered during the translation. In addition, we convert the protein function prediction problem to a language translation problem which is on the new proposed language space, and apply the state-of-the-art Neural Machine Translation Model based on the recurrent neural network to predict protein function. We evaluate the performance of our ProLanGO model and also compare it with another probability based model proposed by us. The result shows that our ProLanGO model (Combination of NMT and extended NMT model) achieves better performance compared to the NMT and extended NMT model on both recall and precision, and similar performance are achieved compared to the probability based model. We also compare with four other state-of-the-art protein function prediction methods, and~our methods show better performance than the SEQ score from SMISS model, but there is still a big gap between our method and the top homologous based method. There are 332 proteins selected by us for the evaluation, but more proteins with real functions would be released later 
, and the official evaluation result from CAFA will also be released. We could rigorously evaluate our method at that time and compare our method with other protein function prediction methods.

In general, we propose a new method for protein function prediction. We compare our method with four state-of-the-art protein function prediction methods. The good performance of our model ProLanGO demonstrates the potential application for protein function prediction problem. There are a lot of improvements that can be done in future. For example, the cross-validation can be included in the training part to improve the reliability of the ProLanGO model. Also, in the Prolan language, we use the frequency of k-mers while the k in the range of three and five is used, the accuracy could be improved by a larger k. In addition, instead of using the frequency of k-mers, a more biologically fragment could be used to divide the protein sequences, such as using a fragment of binding site for a protein. For the bucketing technique, different sizes for each bucket could be tested to additionally improve the performance. The GO terms predicted by our model could be ranked by using other information. Finally, it will be very interesting to see how the order of these k-mers improve the performance.

\subsubsection*{Acknowledgments}

This work was supported by the National Natural Science Foundation of China under grant 61201273 and Natural Sciences Summer Undergraduate Research Program by Pacific Lutheran University.

\subsubsection*{Author contributions}
R.C. and Z.C design and build the system, R.C., C.F., and M.S. carries out the experiments, R.C. builds the server and install programs, R.C, C.F., L.C., M.S., H.J. and Z.C. write the manuscript and give suggestions.

\section*{References}

\medskip

\small

\renewcommand\refname{\vskip -1cm} 

\bibliographystyle{unsrt}

\bibliography{bibtex_entries}

\end{document}